\documentclass[prb,aps,twocolumn]{revtex4}
\usepackage{graphicx}
\input pstricks
\input epsf
\begin{document}
\title{Stacked clusters of polycyclic aromatic hydrocarbon molecules}
\author{M. Rapacioli}
\affiliation{Centre d'\'Etude Spatiale des Rayonnements, CNRS-UPS, 9
av.~du colonel Roche, BP 4346, 31048 Toulouse Cedex 4, France}
\author{F. Calvo}
\affiliation{Laboratoire de Physique Quantique, IRSAMC, Universit\'e
Paul Sabatier, 118 Route de Narbonne, F31062 Toulouse Cedex, France}
\author{F. Spiegelman}
\affiliation{Laboratoire de Physique Quantique, IRSAMC, Universit\'e
Paul Sabatier, 118 Route de Narbonne, F31062 Toulouse Cedex, France}
\author{C. Joblin}
\affiliation{Centre d'\'Etude Spatiale des Rayonnements, CNRS-UPS, 9
av.~du colonel Roche, BP 4346, 31048 Toulouse Cedex 4, France{}}
\author{D. J. Wales}
\affiliation{University Chemical Laboratories, Cambridge CB2 1EW,
United Kingdom}

\begin{abstract}
Clusters of polycyclic aromatic hydrocarbon (PAH) molecules are
modelled using explicit all-atom potentials using a rigid body
approximation. The PAH's considered range from pyrene (C$_{10}$H$_8$)
to circumcoronene (C$_{54}$H$_{18}$), and clusters containing
between 2 and 32 molecules are investigated. In addition to the
usual repulsion-dispersion interactions, electrostatic point-charge
interactions are incorporated, as obtained from density functional theory
calculations. The general electrostatic distribution in neutral or
singly charged PAH's is reproduced well using a fluctuating charges
analysis, which provides an adequate description of the multipolar
distribution. Global optimization is performed using a variety of
methods, including basin-hopping and parallel tempering Monte Carlo.
We find evidence that stacking the PAH molecules generally yields
the most stable motif. A structural transition between one-dimensional
stacks and three-dimensional shapes built from mutiple stacks is observed
at larger sizes, and the threshold for this transition
increases with the size of the
monomer. Larger aggregates seem to evolve toward the packing 
observed for benzene in bulk.
Difficulties met in
optimizing these clusters are analysed in terms of the strong
anisotropy of the molecules. We also discuss segregation in
heterogeneous clusters and vibrational properties in the context
of astrophysical observations.
\end{abstract}
\maketitle

\section{Introduction}
\label{sec:intro}

Polycyclic aromatic hydrocarbons (PAH's) have been proposed as the
carriers of a family of interstellar aromatic infrared bands (AIB's),
observed in many astronomical objects.\cite{allamandolla,leger}
The most intense of these bands, found near 3030, 1610, 1315--1282,
1150 and 885 cm$^{-1}$ respectively, are observed systematically
from different regions of the interstellar medium heated by
starlight. The intensity of these features suggests that PAH's are
the most abundant complex polyatomic molecules in the interstellar
medium, accounting perhaps for as much as 20$\%$ of all carbon in our
galaxy.\cite{Boulanger99} Many laboratory
studies\cite{Szczepanski,Banisaukas, Hudgins94,Hudgins95,Mattioda,Huneycutt,%
Kurtz,Joblin95,Cook,Oomens,Kim} and quantum
chemical calculations or models\cite{deFrees,Pauzat95,Langhoff,Ellinger,Pauzat02,Bauschlicher,Hudgins01,chen99,Pauzat,weingartner}
of single PAH molecules have been performed in
the past, but none has yet provided convincing evidence that single PAH molecules are
actually present in the interstellar medium. This result is likely
due to the specific nature of the
interstellar species, in relation to their formation mechanism
and further processes such as photodissociation. Boulanger {\em
et al.}\cite{boulanger} and Bernard {\em et al.} \cite{bernard}
suggested that the observed free-flying PAH's are produced by
photoevaporation of larger grains. Cesarsky {\em et
al.}\cite{cesarsky} attributed the presence of an infrared continuum
observed in the reflection nebula CED 201 to very small carbonaceous
grains eventually
leading to the AIB's carriers. In recent work, Rapacioli {\em et
al.}\cite{rapacioli} presented strong evidence that these grains are PAH
clusters, and a typical lower size of 400 carbons per cluster was
inferred.

Generally speaking, very little is known about the structure of PAH
clusters. In comparison, assemblies of benzene molecules have received
much more attention.\cite{vandewaal1,vandewaal2,stace,williams,engkvist,%
wallqvist,li,bartell,easter0,easter1,easter2,gonzalez01,khanna}
At empirical levels
of theory, van de Waal\cite{vandewaal1,vandewaal2} and the groups of
Stace,\cite{stace} Bartell\cite{bartell} and Whetten and
Easter\cite{easter0,easter1,easter2} have
investigated assemblies containing up to 13 benzene molecules.
These studies showed a marked preference for the Wefelmeier growth
scheme\cite{wefelmeier} also observed in argon clusters.\cite{hoare}
The (benzene)$_{13}$ cluster, in particular, was shown to exhibit very
stable icosahedral-based structures,\cite{easter2} some of which have
received experimental support. At more sophisticated levels,
electronic structure calculations have also been performed by several
authors.\cite{engkvist,wallqvist,li,gonzalez00,gonzalez01}
The cationic species (C$_6$H$_6$)$_n^+$ have been investigated
experimentally from ion mobility measurements by Rusyniak {\em et
al.}\cite{khanna} These authors also performed structural optimization
using the OPLS force field.\cite{opls}

Naphthalene clusters have been
studied by van de Waal,\cite{vandewaal1} who identified 
patterns similar to benzene clusters. More recently, the structure of the
naphthalene trimer obtained from experimental studies\cite{benharash}
was shown to agree well with one derived from {\em ab initio}
calculations.\cite{gonzalez99}

Anthracene clusters containing up to 5 molecules were investigated theoretically
and experimentally by Piuzzi and coworkers.\cite{piuzzi}
Despite their relatively small size, these clusters seem to show
quite different structures, forming a mainly two-dimensional pattern
in which the long axes of the molecules are aligned, their centers
of mass lying in the same perpendicular plane. Song and
coworkers\cite{song} reported mass spectrometry measurements for anionic
clusters containing up to 16 anthracene molecules.

Heterogeneous clusters made of one or several benzene molecules and
a single naphthalene, anthracene, perylene or tetracene molecule
have been studied, sometimes in the presence of a solvent.\cite{severance,bahatt}
Data for larger PAH molecules is significantly more scarce.
The coronene and circumcoronene dimers were investigated by Miller and
coworkers.\cite{miller84} Aggregates of PAH molecules were
{\em assumed}\/ by Seahra and Duley \cite{duley} to form stacks.
These authors used data from graphite to estimate some translational
vibrational modes of stacked PAH's. However, this assumption did not
employ any atomistic
modelling. Marzec,\cite{marzec} using an extension of the MM2
force field\cite{mm2} and semi-empirical quantum mechanical
approaches, performed local optimizations of stacked structures containing
up to 6 molecules. In this study, the stacks were shown to be stable
for PAH's ranging from dibenzopyrene to dicoronene.\cite{marzec}
Perlstein\cite{perlstein} investigated the transition from
one-dimensional stacks to monolayers and crystal packings of various
molecular species, also using the MM2 classical potential.

While bulk polyaromatic compounds are very difficult to study
experimentally, nearly pure macroscopic samples of hexabenzocoronene
molecules were shown to form a twinned crystal.\cite{goddard95}
Crystallization via epitaxial growth over a graphite surface has also
been reported.\cite{schmitz00} Computer simulations results by Khanna
and coworkers\cite{khanna04} indicate that large sets of coronene or
circumcoronene molecules do indeed crystallise and melt through a
first-order process without exhibiting intermediate liquid crystal
behavior. More generally, the bulk phases of PAH molecules can be
expected to show similar phase diagrams as the model discotic particles
studied by several authors.\cite{bates,zewdie,caprion}

Our present interest lies in the size range between the very small and
very large sizes, for aggregates containing about 1000 carbon atoms.
We wish to address the following questions:
\begin{itemize}
\item[(i)] Are the stacks of PAH molecules actually the most stable
form for clusters?
\item[(ii)] How large can these stacks be as a function of the PAH monomer
size?
\item[(iii)] How does the cluster structure evolve toward the bulk
morphology?
\end{itemize}
In addition to these main concerns, we intend to address not only
clusters of the same PAH molecule, but also some heterogeneous
assemblies, as the experimental situation we are referring to does not
have restrictions on the variety of PAH's involved.

Locating the most stable structures of molecular clusters can be quite
difficult.\cite{bhwater,hartke,kabrede} For example, water clusters have
been found to exhibit a multi-funnel potential energy
surface,\cite{bhwater,walesnat,walesbook} where numerous low energy
minima exist, but can only be interconnected by long pathways.
Global optimization is significantly harder for rigid body water
clusters than for most atomic clusters with a comparable number
of degrees of freedom.\cite{bhwater}
PAH clusters, which are made of very anisotropic monomers, represent a novel
challenge for molecular optimization, somewhere in between the difficulties
associated with atomic clusters and those of biological
molecules.\cite{walesscheraga}

In the present work, we have attempted to locate the stable structures
of assemblies of relatively large PAH molecules ranging from pyrene to
circumcoronene. We combine results from electronic structure
calculations and atomistic modelling to describe the intermolecular
interactions, within a rigid body approximation for the PAH
molecules. Global optimization using the basin-hopping\cite{mcmin,bh}
and parallel tempering\cite{ptmc} Monte Carlo methods then
provide estimates of the most stable isomers, for clusters containing
up to 32 molecules.

The paper is organized as follows. In the next section, we give the
basic description of our model, as well as some details about the
electronic structure calculations and the optimization
procedure. Section \ref{sec:struct} presents our results for the
structures, starting with the dimers. A more detailed analysis is provided
in the case of coronene clusters, and the relative stability of some
important structural motifs is investigated. Sec. \ref{sec:struct}
ends with some results for heterogeneous clusters. In
Sec. \ref{sec:vib} we briefly discuss intermolecular vibrational
modes. We comment in Section \ref{sec:disc} on the difficulties
encountered in our global optimization approach and in the robustness
of our results with respect to the quality of the atomistic
modelling. Finally, we summarize and discuss the astrophysical relevance
of our results in Sec. \ref{sec:ccl}.

\section{Methods}
\label{sec:methods}

Our goal here is to provide good candidates for the most stable
structures of clusters of PAH molecules. These structures are expected
to be relevant at very low temperatures, where intramolecular
vibrations are unlikely to be excited. This legitimises our main
approximation that the PAH molecules can be treated as rigid bodies.

\subsection{Intermolecular potentials}

The PAH molecules considered are pyrene (C$_{16}$H$_{10}$),
coronene (C$_{24}$H$_{12}$), ovalene (C$_{30}$H$_{16}$),
hexabenzocoronene (HBC, C$_{42}$H$_{18}$), octabenzocoronene
(OBC, C$_{46}$H$_{18}$), and circumcoronene (C$_{54}$H$_{18}$). The
cluster sizes range from 2 to 32 depending on the monomer, so we will
be dealing with a rather large number of atoms. Considering the known
difficulty of global optimization,\cite{walesscheraga} it seems
obvious that the atomistic modelling should be limited to a moderate
numerical cost---but retaining the important chemical
features. In particular, the multipolar electrostatic description
employed in Ref.~\onlinecite{piuzzi} is not appropriate for an
initial survey, although such ideas could provide useful
corrections. Following previous efforts on benzene clusters by
several groups,\cite{vandewaal1,stace,bartell,easter1,easter2}
we have chosen to describe the intermolecular potential in PAH
clusters as the result of two contributions:
\begin{equation}
V=\sum_{i<j} \sum_{\alpha\in i} \sum_{\beta \in j}
\left[ V_{\rm LJ}({\bf r}_{i_\alpha,j_\beta}) + V_{\rm Q}({\bf
r}_{i_\alpha,j_\beta})\right],
\label{eq:vpot}
\end{equation}
where $V_{\rm LJ}({\bf r}_{i_\alpha,j_\beta})$ denotes the
repulsion-dispersion energy between atom $i_\alpha$ of
molecule $i$ and $j_\beta$ of molecule $j$, and $V_{\rm Q}$ the
electrostatic interaction between the partial charges carried by the
molecules. Here and in the following, the PAH molecules are denoted by
roman letters, and greek letters are used for the atoms of each molecule.

The Lennard-Jones (LJ) form is used for the repulsion-dispersion, with
parameters $(\sigma_{\rm CC},\sigma_{\rm HH},\sigma_{\rm CH},
\varepsilon_{\rm CC},\varepsilon_{\rm HH},\varepsilon_{\rm CH})$ taken
from Ref.~\onlinecite{vandewaal1}. The electrostatic contributions are
based on the point charges $\{q_i\}$ located on the atom sites.

\subsection{DFT calculations and electrostatic modelling}

The molecular geometries and partial charges were initially determined
from electronic structure calculations on PAH monomers. 
The molecular geometries were those determined in tight-binding calculations. 
The  partial atomic charges were initially determined from electronic
structure calculations using the B3LYP hybrid functional and
the {\sc Gaussian03} package.\cite{gaussian} Those
calculations were carried out using core pseudopotentials on carbon
as determined by Durand and Barthelat,\cite{durand}
and tabulated in semi-local form by Bouteiller {\em et al.}\cite{bouteiller}
The gaussian type orbital (GTO) basis set used was [4s/4p/1d]
contracted into [31/31/1] on
the carbon atoms and [4s/1p] contracted into [31/1] on the hydrogen
atoms. The basis sets are specified fully in table \ref{tab:basis}.

In addition to the neutral PAH molecules, calculations were also
performed on the singly charged anions and cations at the neutral
geometry. These results allow us to investigate the relative stability of
zwitterionic species.

\begin{table}[htb]
\begin{tabular}{|c|cc|cc|cc|}
\hline
   & \multicolumn{2}{c|}{$s$} & \multicolumn{2}{c|}{$p$} &
\multicolumn{2}{c|}{$d$}     \\     
   & Exp. & Cont. & Exp. & Cont. & Exp. & Cont. \\
\hline
   & 2.382013 & -0.242140&8.609570 & 0.043653 & 0.80000 &  1.0\\
\,~\rput(0,-.25em){C}~ & 1.443065 &  0.185265&1.943550 & 0.209497 & &\\
   & 0.405847 & 0.591283 &0.542798 & 0.502761 && \\
   & 0.138427 & 1.0      & 0.152496& 1.0      &&\\
\hline
   &13.24876&  0.019255 &  0.90000 & 1.0   &&\\
\,~\rput(0,-.25em){H}~ &2.003127&  0.134420 & & &&\\
   &0.455867&  0.469565 &  &  &&\\
   &0.124795&  1.0 & & && \\
\hline
\end{tabular}
\caption{Atomic GTO basis sets employed for carbon and hydrogen.}
\label{tab:basis}
\end{table}

The extraction of point atomic charges from {\em ab initio} calculations 
is rather problematic since charges are not uniquely defined. The
well-known Mulliken definition\cite{mulliken} faces the overlap problem
and the corresponding results may depend strongly on the basis set. 
Other definitions
of atomic charges can be found, such as the natural bond orbital (NBO)
definition\cite{carpenter} or the Bader definition,\cite{bader} in
which the charges are determined from the topology associated with the
density. For the present purpose however, since the charges are
essentially used to extract an intermolecular potential rather than
describing the intramolecular bonding, we
choose to compute the charges which best fits the actual electrostatic
potential on the van der Waals surface associated with each molecule.
This choice is similar to the strategy followed by Piuzzi and
coworkers,\cite{piuzzi} which provided reasonable results for
the anthracene dimer by consistently reproducing the {\em ab initio} 
interaction energies for various relative orientations of the molecules.

The fitted charges were determined here using the {\sc Gaussian03}
package for all the species mentioned above. As expected, the
electrostatic potential fitted (EPF) charges are significantly different
from those given by the Mulliken analysis. They show more continuity
from one PAH to one another, especially in terms of the populations
on hydrogens with respect to carbon atoms. With the present basis set,
the Mulliken analysis often leads to negative charges on hydrogen atoms,
whereas both NBO and EPF charges for neutral PAH's are always positive
on hydrogen. Table \ref{tab:charge} illustrates the values
obtained for coronene. The other PAH's behave similarly, except that
they show larger fluctuations within a given molecule, especially when
the symmetry is lower.

\begin{table}[htb]
\begin{tabular}{|c|r|r|r|}
\hline
 shell &  Mulliken & NBO      & EPF \\
\hline
C1(6)     &  0.0040  &  $-0.0082$ & 0.0004\\
C2(6)     &  $-0.0295$  & $-0.0492$ & 0.0799\\
C3(12)     &  0.0299   & $-0.1925$ & $-0.1697$\\
H(12)      &  $-0.0172$  &  0.2212 &  0.1295\\
\hline
\end{tabular}
\bigskip
\caption{Mulliken, natural bond orbital, and electrostatic
potential fitted charges for carbon and
hydrogen atoms of coronene. The carbon atoms are labelled according to
their radial location from central to peripheral, and the number of
equivalent atoms is also indicated in parentheses.}
\label{tab:charge}
\end{table}

After the charges were determined, a simple electrostatic model was
constructed to rationalize the DFT results. The model is similar to the
charge equilibration scheme of Mortier {\em et al.},\cite{mortier} and
also to the fluctuating charges model of Rapp\'e and
Goddard.\cite{goddard} Interestingly, some links between DFT
and fluctuating charges potentials have been
established.\cite{fqdft} Briefly, at a given molecular geometry
the atomic charges $\{ q_i \}$ are defined
in order to minimize the global electrostatic energy under the
constraint of total charge conservation:
\begin{equation}
E_Q = \sum_i \left( \varepsilon_i + \frac{1}{2}U_i q_i^2 \right)
+ \sum_{i<j} J_{ij} q_i q_j + \lambda \left( \sum_i q_i - Q_0 \right).
\label{eq:flucq}
\end{equation}
In the above equation, $\varepsilon_i$ and $U_i$ are the
electronegativity and hardness of atom $i$, respectively, and
$J_{ij} = (r_{ij}^3+\gamma_{ij}^{-3})^{-1/3}$. The Lagrange multiplier
$\lambda$ ensures that the total charge is constant and equal to $Q_0$.
This electrostatic model has only 4 parameters, namely
$\varepsilon_{\rm C}-\varepsilon_{\rm H}$, $U_{\rm C} = \gamma_{\rm
CC}$, $U_{\rm H}=\gamma_{\rm HH}$, and $\gamma_{\rm CH}$. These
parameters were optimized to reproduce the partial charges of the
neutral coronene molecule, as obtained from density functional
calculations. We found the charges carried by the atoms of all other PAH
molecules to be very satisfactorily reproduced by this model, always
within 5\%, not only for neutrals, but also for anionic or cationic
PAH's. Therefore, the complete electrostatic atomistic description is
gathered into the parameters of this simple model, which read
$\varepsilon_{\rm C}-\varepsilon_{\rm H} = 1.86$~eV, $U_{\rm
C}=0.98$~eV, $U_{\rm H}=1.65$~eV, and $\gamma_{\rm CH}=2.11$~eV.

Before going on, it is important to notice that the present
fluctuating charges model not only accounts for (first-order)
Coulombic interactions. The charges obtained are intrinsically the
solution of a linear many-body equation, and the resulting energy
incorporates polarization effects. In addition to providing all the
important electrostatic characteristics of the single molecule, the
fluctuating charges model can be extended to treat an assembly of PAH
molecules. For this purpose, Eq.~(\ref{eq:flucq}) must be extended to
include interactions between all the atoms, and the conservation of
total charge over each molecule is achieved through the introduction of a
corresponding number of Lagrange multipliers.\cite{berne} In principle, charge
transfer among a limited or complete set of molecules could also be
investigated, but we have not considered this possibility here.

Allowing charges to fluctuate within a PAH cluster has a very heavy
computational cost, which turned out to be incompatible with performing
large scale optimization. We therefore restricted the model to fixed
charges. However, for some of the typical equilibrium structures we
found, we checked that the assembly of several PAH molecules was
consistent with the fixed charges approximation. To do this we computed
the average $\langle | q_{i\alpha} - q_{i\alpha}^{(0)}|\rangle$ over
all atoms and all molecules from the charges $\{ q_{i\alpha}^{(0)} \}$
in the single molecules and their values $\{ q_{i\alpha} \}$ in the
relaxed cluster. For all clusters investigated, the average charge always
remained lower than 0.1\%. Obviously, at finite
temperatures intramolecular vibrations are likely to induce
significant variations in the effective atomic charges, and the fixed
charges approximation will become less accurate.

\subsection{Global optimization}

We attempted to locate good candidates for the global minima of PAH clusters
using several methods. The basin-hopping (BH), or Monte
Carlo plus minimization method\cite{mcmin,bh} was used, especially for the
smaller clusters. This unbiased algorithm performs a random walk on
the transformed potential energy surface, which consists only of the
local minima. This walk requires quenches to be carried out after each
molecular displacement. Here, a displacement means that all molecules
are rotated and translated randomly by some (rather large) amount, at
the same time.

The basin-hopping method was previously used for a variety of other
atomic\cite{bh,bhmetal} and molecular\cite{bhwater,bhn2,bhcapsid}
clusters. However, we found it rather inefficient for
moderately large PAH clusters, for reasons that will be discussed below.
Therefore, we also employed parallel tempering Monte Carlo\cite{ptmc}
as an alternative optimization tool. Even though this method was
originally designed for reducing broken ergodicity in glassy systems,
it has since been used successfully in global optimization
problems.\cite{deem,fuchs} It essentially consists in performing several MC
trajectories at increasing temperatures, and allowing exchange moves
between adjacent trajectories in addition to the standard molecular
displacements. In practice, we chose 20 temperatures in the range
1~K$\leq T\leq$80~K. Periodic quenches from
some trajectories were also considered to provide extra
isomers. Finally, many low energy structures were found by
construction from simple motifs, using the structures obtained from unbiased methods
for guidance.

\section{Structures}
\label{sec:struct}

The interaction between molecules is essentially additive within our
model. However, due to the strong aspherical character of the
molecules, this interaction is also expected to be highly anisotropic.
In view of the numerous previous results for dimers of PAH's, we first
focus on these species as a first step toward larger aggregates.

\subsection{Dimers}

\begin{figure}[htb]
\setlength{\epsfxsize}{8cm}
\leavevmode\centerline{\epsffile{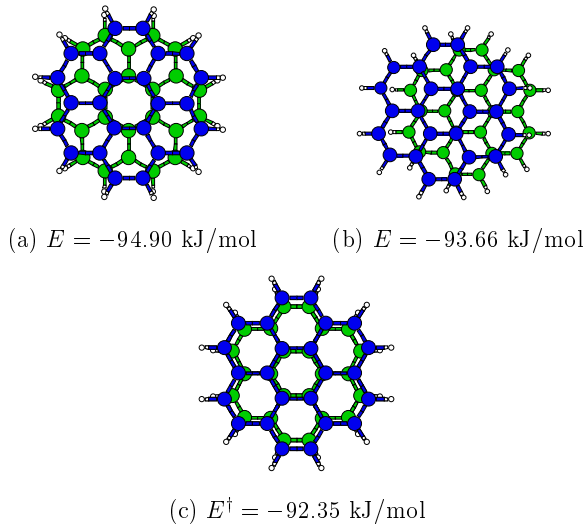}}
\caption{Low energy structures of the coronene dimer. (a) Twisted
stack; (b) parallel-displaced stack; (c) superimposed stack (saddle point).}
\label{fig:corodimer}
\end{figure}

The $D_{6h}$ coronene molecule provides a convenient example of large
PAH's, especially from the computational point of view. The most stable
structures of its dimer are shown in Fig.~\ref{fig:corodimer}, (a) and (b).
These structures are perfect stacks, in the sense that the sixfold
axes and molecular planes are parallel.
A detailed analysis of the various contributions to the binding
energy shows that the repulsion-dispersion between carbon atoms
dominates. Therefore it may be energetically favorable to shift the
molecules along some parallel displacement, or to rotate them
around their common sixfold axis. The resulting structures show an
increase in the number of nearest neighbors, but the parallel
displaced geometry is slightly penalized at its edges due to fewer
van der Waals bonds involving hydrogen atoms. Interestingly, the
perfectly superimposed stack, Fig~\ref{fig:corodimer}c, is not a
true minimum but a saddle point for our potential. This result contrasts
with that obtained by Marzec,\cite{marzec} who found the face-to-face
configuration to be the lowest in energy. It should be noted here that
the charges used by this author, obtained from the MINDO3 and ZINDO1
semi-empirical quantum mechanical calculations, are significantly lower
than ours (always between $10^{-3}$ and $10^{-2}$). However, even after
removing the electrostatic contribution from our calculation we still
find that this conformation is a saddle. This result may indicate that
the simple steepest-descent local minimization performed in
Ref.~\onlinecite{marzec} did not fully converge.

\begin{figure}[htb]
\setlength{\epsfxsize}{8cm}
\leavevmode\centerline{\epsffile{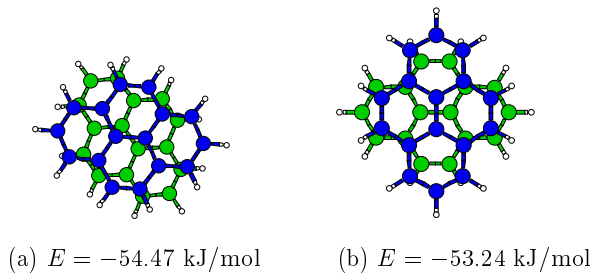}}
\caption{Low energy structures of the pyrene dimer. The angle between
the long axes is (a) 50$^\circ$ and (b) 90$^\circ$.}
\label{fig:pyredimer}
\end{figure}
A pair of pyrene molecules also exhibits stacks, as shown in
Fig.~\ref{fig:pyredimer}. The superimposed structure analogous to (c) of
Fig.~\ref{fig:corodimer} spontaneously transforms into a twisted stack where
the angle between the long axes is about 30$^\circ$. Another stable
stack is also found with perpendicular long axes. This isomer, (b) in
Fig.~\ref{fig:pyredimer}, will be denoted as `staggered' below.

Smaller aromatic molecules also display stable stacked
dimers, even though T-shaped geometries are found competitive for
benzene, naphthalene and anthracene.\cite{gonzalez00,piuzzi,severance}
(Benzene)$_2$ is slightly more stable when T-shaped, but increasing
the number of aromatic rings favors stacked shapes.\cite{severance}
In both (naphthalene)$_2$ and
(anthracene)$_2$, the crossed stacks are marginally preferred to
the parallel displaced stacks.\cite{gonzalez00,piuzzi}
However, we note that, among T-shaped conformations, 
those with parallel long axes show an enhanced stability.\cite{piuzzi}

The trends seen in the naphthalene and anthracene dimers are also
observed in the present work for (coronene)$_2$ and (pyrene)$_2$.
In the latter case, the parallel displaced isomer leads to structure
(a) in Fig.~\ref{fig:pyredimer} after minimization.
In the circumcoronene dimer, not shown, the crossed stacked
structure is also the most stable, the parallel displaced stack being
1.7~kJ/mol higher in energy. Again, the perfectly superimposed dimer
is not a stable minimum.

The favored dimer structure results from a balance between maximizing
the number of weak C--C intermolecular contacts, and minimizing the Coulomb repulsion
between hydrogen atoms. The gain in repulsion-dispersion energy
is larger than the Coulomb penalty in the crossed stack relative to
the parallel displaced stack. We have also investigated
cohesion in zwitterionic (PAH$^+$PAH$^-$) stacks. For all the
clusters investigated, the crossed stacks were further stabilized
with respect to the parallel-displaced isomer. In these cases, the
Coulomb interaction becomes strongly attractive instead of being slightly
repulsive for the neutral forms.

>From the DFT calculated vertical ionisation potentials and electron
affinities, and the binding
energy of the ion pair, the energy of the zwitterion can be estimated
relative to the neutrals. The binding energy was calculated using
the same intermolecular potential, with fixed partial charges
corresponding to each particular ion. This scheme neglects the
resonance interaction PAH$^+$PAH$^-\rightleftharpoons$PAH$^-$PAH$^+$.
We find the relative zwitterion energies $\Delta E=4.23$, 3.38, and
2.87~eV above the the neutral minimum for the pyrene, coronene,
and OBC dimers, respectively. These values are similar to the result
reported by Piuzzi {\em et al.} for anthracene trimers of around 3.5~eV.\cite{piuzzi}

Finally, the distance between molecular
planes was found to lie between 3.56\,\AA\ for (pyrene)$_2$ and 3.50\,\AA\ for
(circumcoronene)$_2$. These values compare well with the data reported
by Marzec,\cite{marzec} as well as the experimental value for crystal
packings obtained by Goddard {\em et al.} on HBC.\cite{goddard95}

\subsection{Coronene clusters}

\begin{figure}[htb]
\setlength{\epsfxsize}{9cm}
\leavevmode\centerline{\epsffile{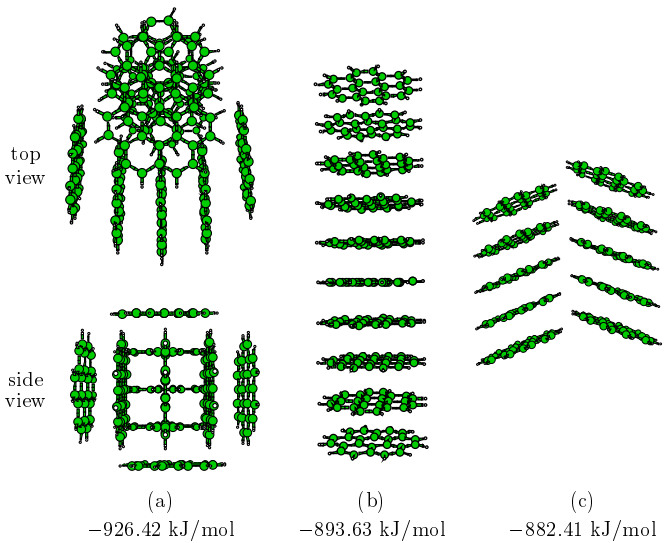}}
\caption{Lowest energy structures of the (coronene)$_{10}$ cluster.}
\label{fig:coro10}
\end{figure}

The lowest-energy structures of coronene aggregates are found
to be single stacks as long as the number of molecules does not exceed eight.
In these stacks, the orientations alternate so that adjacent molecules
are staggered.  Single stacks remain local minima for
larger numbers of molecules, mainly due to the short-range character of
the interaction. However, above eight molecules lower energy structures appear, as
illustrated in Fig.~\ref{fig:coro10} for (coronene)$_{10}$.

The most stable conformation is shown in Fig.~\ref{fig:coro10}a. In
this global minimum the two `C'-shaped stacks adopt a configuration
that looks rather like a handshake. The tilted stacks of
Fig.~\ref{fig:coro10}c, which we expected to be natural evolution
of the one-dimensional stack, turn out to be less favorable
at this size. The special stability of the `handshake' structure is
due to its compactness, while conserving the primary stack motifs as
basic units. Not surprisingly, other combinations of stacks such as
4+6 or 3+7 are higher in energy than the 5+5 structure of
Fig.~\ref{fig:coro10}a.

Building upon the (coronene)$_{10}$ cluster, extra molecules first
tend to add at the tips of existing stacks. However, the `handshake'
8+8 stacks are actually less stable than a 8+4+4 triple stack
where the two short 4-stacks make opposite turns around an S-shaped
8-stack. In comparison, the 6+5+5 triple stack built by adding a
third column to the structure of Fig.~\ref{fig:coro10}c in a
triangular fashion is slightly higher in energy. This difference
is mainly due to the geometrical frustration arising when trying to
alternate the new molecular planes with those already present.

\begin{figure}[htb]
\setlength{\epsfxsize}{8cm}
\leavevmode\centerline{\epsffile{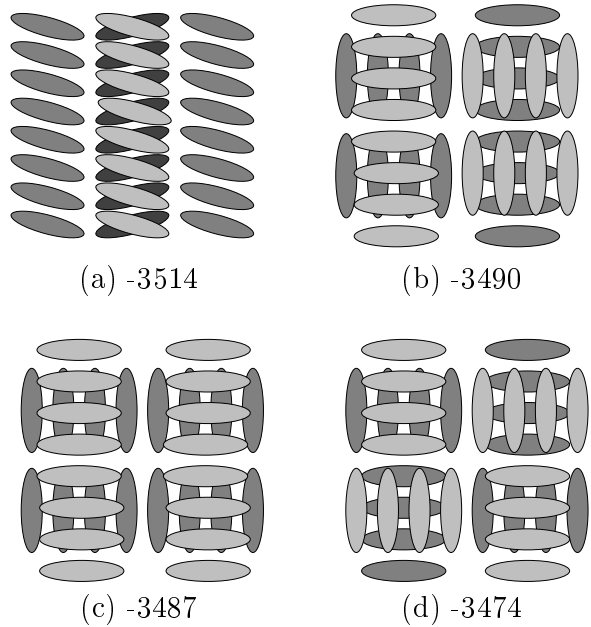}}
\caption{Lowest energy stacked structures of the (coronene)$_{32}$ cluster. The
total binding energies are given in kJ/mol.}
\label{fig:coro32}
\end{figure}

We attempted to grow the S-shaped structure further by adding molecules
at the six extremities, but this proved not to be the optimal route for
larger clusters. There are many ways to put together 4-stacks in order
to generate a (PAH)$_{32}$ cluster. The four most stable such conformations
are sketched in Fig.~\ref{fig:coro32}. 

Among these conformations, three schemes based on alternating stacks
oriented perpendicularly are
very close in energy, independent of the stack sizes. However, four
8-stack, tilted columns with parallel axes provide the most stable
conformation at this size. This structure, depicted in
Fig.~\ref{fig:coro32}a, appears as a clear precursor to the herringbone
crystal packing reported by Khanna {\em et al.}\cite{khanna04b} from
computer experiments. Even though we did not carry out a systematic search
for minima above 20 PAH molecules, decreasing the stack size from the
(coronene)$_{32}$ herringbone motif of Fig.~\ref{fig:coro32}a remained
the most stable structure down to less than 24 molecules. Below this size,
perpendicular stacks similar to Fig.~\ref{fig:coro32}c become the
global minimum. The fully intertwinned structure exemplified in
Fig.~\ref{fig:coro32}d is only optimal at sizes lower than 20.
Therefore, it seems that bulk structure appears in coronene at about
$N\sim 24$ molecules.

The appearance of bulk features in the structural properties of clusters
has attracted much attention in the past, especially in molecular
clusters.\cite{fuchsn2,fuchsco2,calvon2} It was noticed in particular that
more isotropic species such as argon or nitrogen exhibit the bulk crystalline
character at much larger sizes than anisotropic molecules like
CO$_2$.\cite{fuchsco2} The results for coronene
thus seem to confirm this observation.

\subsection{Stability}

\begin{figure}[htb]
\setlength{\epsfxsize}{8cm}
\leavevmode\centerline{\epsffile{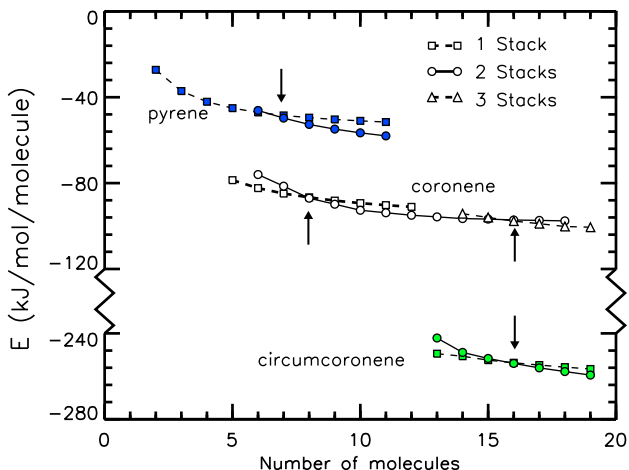}}
\caption{Binding energy of single- and multiple-stack structures of 
pyrene, coronene, and circumcoronene assemblies.}
\label{fig:12stacks}
\end{figure}

The previous results obtained for coronene clusters generally hold for
other PAH molecular units. The relative stability of single-
or multiple-stack structures is shown in Fig.~\ref{fig:12stacks}.
Pyrene clusters form single stacks,
then `handshake' stacks above seven molecules. In circumcoronene clusters, the
one-dimensional stack remains the global minimum until 17 molecules are
reached. The single stacks remain stable longer for larger PAH's, probably
because this motif is favorable until the length exceeds the diameter
of the molecule itself. From $N=18$ and above, circumcoronene clusters
favor the herringbone conformation pictured in Fig.~\ref{fig:coro10}c for
coronene. This result again confirms the idea that increasingly anisotropic
molecules show earlier signatures of bulk structure.

\begin{figure}[htb]
\setlength{\epsfxsize}{8cm}
\leavevmode\centerline{\epsffile{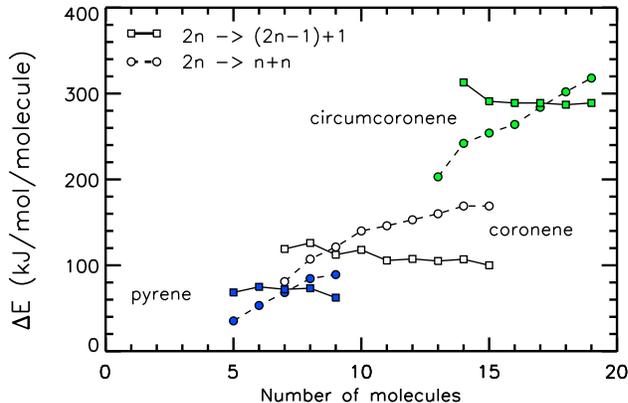}}
\caption{Dissociation energy of two-stack PAH assemblies: removal of one
molecule versus dissociation between the two stacks.}
\label{fig:dissoc}
\end{figure}

The relative stability of stacked configurations is also addressed in
Fig.~\ref{fig:dissoc} where we compare the dissociation energy of
double stack structures (herringbone or `handshake' conformations)
into their two stacks with that of the most weakly bound molecule.
The latter molecule is located at one of the four extremities of the stacks.
Again, we consider pyrene, coronene and circumcoronene clusters.
Obviously, the dissociation energy of a double stack into two
stacks grows approximately linearly with stack size. Conversely, the
binding energy of one single molecule remains nearly constant, and
depends mainly on the monomer. These
trends, evident in Fig.~\ref{fig:dissoc},
explain why it becomes more interesting to remove a single molecule
only after some crossover size, below which the extra stability of the
small stacks dominates. The crossover sizes, seven for pyrene, nine for
coronene and 17 for circumcoronene, correlate with the
change in stability of double stacks with respect to single stacks.
This result emphasizes that, unlike most weakly bound clusters, monomer
dissociation may not always be the lowest energy dissociation pathway for such species.

\subsection{Demixing in heterogeneous clusters}

To date, no specific PAH has been clearly identified as 
the carrier of the interstellar AIB's. In fact, 
these bands probably result from the emission of a population 
of several types of PAH molecules,\cite{schutte,pech} and we
expect interstellar PAH
assemblies to be quite heterogeneous. It is therefore important
to characterize the effects of mixing on the most stable
conformations. Table \ref{tab:mix4} shows the relative stability
of the various possible stacks containing two pairs of identical
molecules based on sixfold symmetry. As in homogeneous
clusters, we were not able to locate structures more stable than
these perfect stacks. Stack (a), in which the larger PAH's occupy
interior sites, maximizes the number of nearest neighbors between
carbon atoms and is energetically the most favorable. Similar arguments
explain the ordering of isomers (a--d). As expected, the
relative stability increases with the disparity
between the PAH molecules: the energy of stack (b) is smaller than
those of stack (a) by about 10\% for (coronene+circumcoronene)$_2$,
but only 8.7\% for (coronene+HBC)$_2$ and 3.4\% for (HBC+circumcoronene)$_2$.
\begin{table}
  \begin{tabular}{|c|c|c|c|c|}
    \hline
\vbox to 1cm{} stack type    &
    \includegraphics[width=1cm]{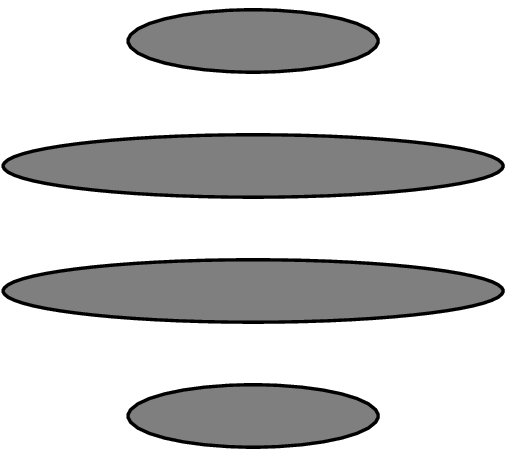}&
    \includegraphics[width=1cm]{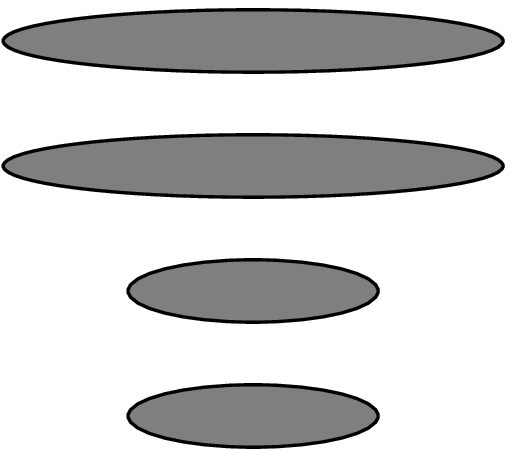}&
    \includegraphics[width=1cm]{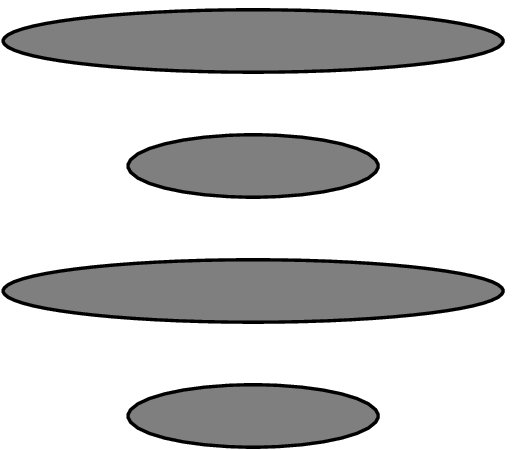}&
    \includegraphics[width=1cm]{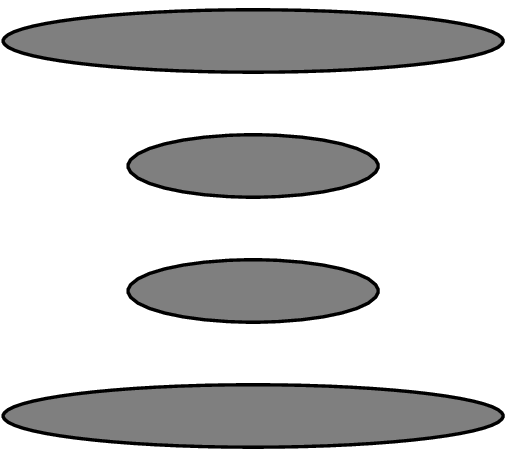}\\
    & (a)  & (b) & (c) & (d)\\
    \hline
  (coronene)$_{2}$(HBC)$_{2}$& $-$473.3 & $-$435.5 & $-$515.5 & $-$376.4\\
  \hline
  (coronene)$_{2}$(circ.)$_{2}$& $-$572.8 & $-$518.7 & $-$478.0 & $-$418.4\\
  \hline
  (HBC)$_{2}$(circ.)$_{2}$& $-$722.3 & $-$695.9  & $-$687.4  & $-$663.1\\
  \hline
  \end{tabular}
  \caption{Binding energies (kJ/mol) of the lowest minima obtained 
for clusters composed of two pairs of PAH molecules. Circ.~refers
to the circumcoronene molecule.}
\label{tab:mix4}
\end{table}

This segregation property also appears for larger clusters.
As an example we have investigated the heterogeneous assemblies
containing 6 large and 4 small PAH molecules of the coronene, HBC,
and circumcoronene types. The putative global minima we found are all
based on the same structures obtained for the homogeneous coronene cluster
depicted in Fig.~\ref{fig:coro10}. For each of these isomers, the
optimal locations of the smaller PAH are at the two or four
tips of the stacks. This result is again in agreement with the energetic
arguments used to explain the trends in Table~\ref{tab:mix4}.

In Table~\ref{tab:mix46} we compare the relative energies
of these three conformations for the three mixtures.
\begin{table}
  \begin{tabular}{|c|c|c|c|}
    \hline
conformation    & intertwinned & single stack & parallel \\
type & stacks (a) & (b) & stacks (c) \\
    \hline
    (coronene)$_{4}$(HBC)$_{6}$ & $-$1430.1 & $-$1500.7 & $-$1442.2 \\
    \hline
    (coronene)$_{4}$(circ.)$_{6}$ & $-$1725.7 & $-$1826.2 & $-$1749.9 \\
    \hline
    (HBC)$_{4}$(circ.)$_{6}$ & $-$2058.4 & $-$2176.2 & $-$2069.2 \\
    \hline
  \end{tabular}
  \caption{Binding energies (kJ/mol) of the lowest minima
obtained for clusters of six large and four smaller PAH molecules. (a), (b)
and (c) refer to the labels of Fig.~\protect\ref{fig:coro10}, while
circ.~refers to circumcoronene.}
\label{tab:mix46}
\end{table}
Substituting the six innermost molecules of the single stack (b) always
yields the optimal structure. This result reflects the extra stability
of single stacks of large molecules, as seen previously. Hence
the geometry is essentially determined by the larger PAH molecule.
This observation cannot easily be generalized to arbitrary numbers
or small and large molecules, and it would be interesting to look
further into the effects of changing composition on the relative
stability of heterogeneous assemblies.

\section{Intermolecular vibrations}
\label{sec:vib}

Our eventual interest lies in interpreting experimental infrared spectra
by identifying the possible PAH assemblies responsible. Although the present
paper is mainly devoted to the prediction of structure, we wish to address here the
intermolecular vibrations of the smallest clusters (dimers), leaving a more
general discussion about larger assemblies and the couplings with
intramolecular motion to future work.

The normal mode frequencies were obtained by double numerical
differentiation of the potential energy and appropriate mass weighting.
The main vibrational modes are characterized
in Table~\ref{tab:vib} for all dimers of the same PAH ranging from pyrene
to circumcoronene, as well as the zwitterionic form of (coronene)$_2$ for
comparison.
\begin{table}
  \begin{tabular}{|c|c|c|c|c|}
    \hline
\vbox to 1cm{} mode (cm$^{-1}$)  &
    \includegraphics[width=1.2cm]{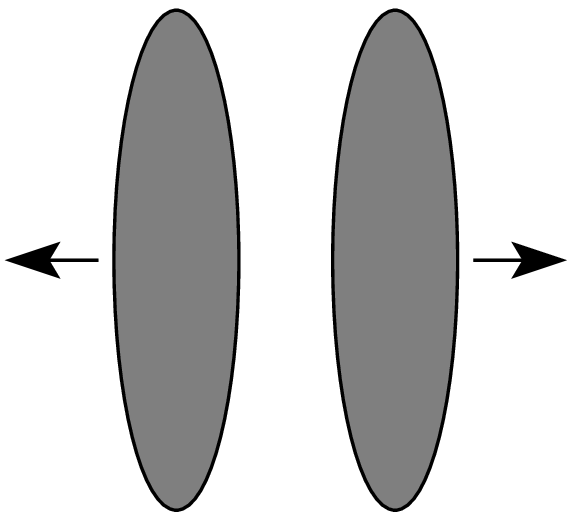}&
    \includegraphics[width=1cm]{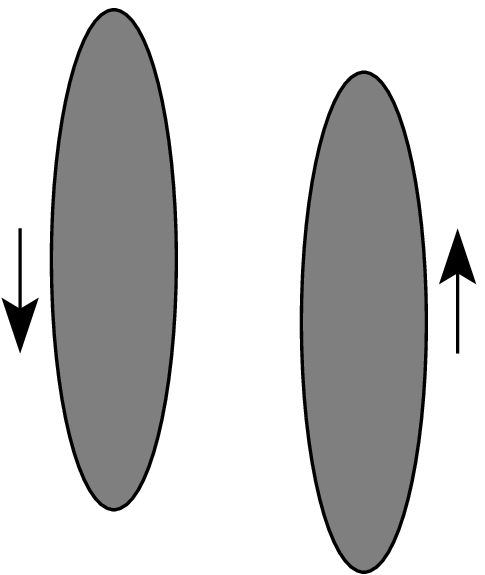}&
    \includegraphics[width=.7cm]{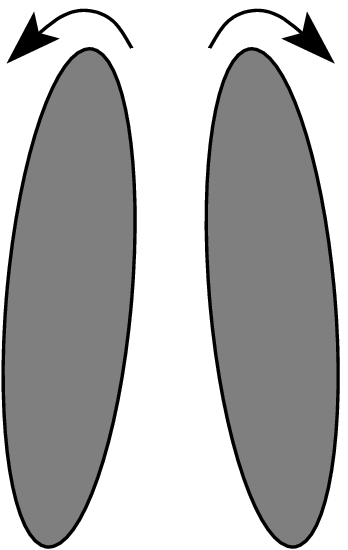}&
    \includegraphics[width=1cm]{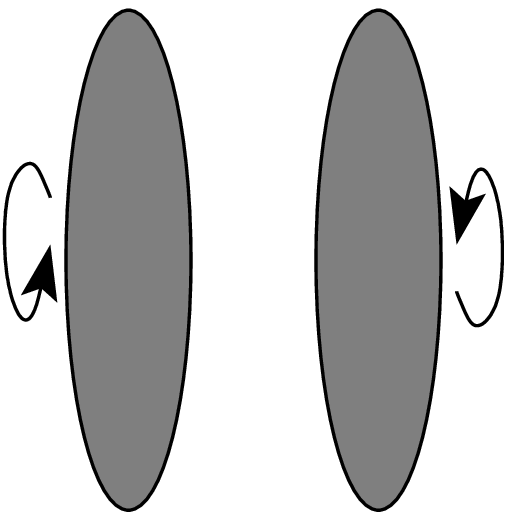}\\
    & (a)  & (b) & (c) & (d)\\
    \hline
    (pyrene)$_{2}$        & 55.1 & 5.5 & 55.5 & 3.6 \\
    \hline
    (coronene)$_{2}$      & 58.9 & 4.3 & 55.9 & 5.6 \\
    \hline
    (ovalene)$_{2}$       & 59.9 & 6.6 & 69.6 & 5.0 \\
    \hline
    (HBC)$_{2}$           & 62.1 & 4.8 & 58.1 & 3.5 \\
    \hline
    (OBC)$_{2}$           & 62.1 & 4.9 & 64.5 & 2.3 \\
    \hline
    (circ.)$_{2}$         & 62.9 & 4.1 & 61.0 & 3.1 \\
    \hline
  coronene$^+$coronene$^-$& 69.0 & 6.8 & 81.3 & 47.4 \\
    \hline
  \end{tabular}
  \caption{Frequencies (cm$^{-1}$) for the vibrational modes of several
PAH dimers. For the shearing (b) and bending (c) modes, an average of
the two values is given.}
\label{tab:vib}
\end{table}
As expected from the strong anisotropy of the molecular interactions,
the vibrational modes can generally be gathered into two separate groups,
the bending and breathing motions being the stiffest and the shearing and
twisting motions the softest. The frequency of the breathing mode
grows quite regularly with increasing PAH mass. For the circumcoronene
dimer, this frequency is quite close to the value generally accepted for
graphene layers, namely $f_0=63.7$~cm$^{-1}$. This result indicates that our
atomistic modelling is physically correct.

The increase of the frequencies for larger PAH's is a natural
consequence of the
increasing strength of the repulsion-dispersion interactions. Both the
repulsive and attractive parts of the potential increase roughly linearly
with the surface area of the PAH molecules, because they are both
short-ranged. At first order, the force
constant scales approximately linearly with the PAH mass, therefore the
associated frequency remains constant. However, the attractive part
of the potential has a shorter range than its repulsive part, so
it increases somewhat faster with the mass.
Thus the effective force constant increases slightly more rapidly than
linearly with the mass, in agreement with the observed increase in
the frequency. This effect is magnified by the decreasing role of the
Coulomb forces, which become negligible in the limit of two graphene sheets.

The bending modes generally show similar frequencies to the breathing mode,
and also the same pattern. However, less symmetrical molecules, such as ovalene
or OBC, have non-degenerate bending modes, resulting in smaller average
values. Both the twisting and shearing modes are comparatively very soft,
reflecting the small corrugation of the molecular planes.

Lastly, the vibrations of the zwitterionic form of the coronene dimer,
chosen as a typical example, display significantly stiffer modes.
In this system, the Coulomb interactions become dominant at long
range, which affects the shearing, and especially the twisting modes.

Larger assemblies of PAH molecules show a greater variety of modes. Stack
vibrations have been considered by Seahra and Duley\cite{duley} in the
one-dimensional approximation of longitudinal modes only. However, the
results obtained in the previous section show that multiple-stack
conformations should become more stable as more and more PAH molecules
are added. This result calls for a more complete study of vibrational
properties, including the possible influence of intramolecular motion.

\section{Discussion}
\label{sec:disc}

\subsection{Difficulties with optimization}

In the original basin-hopping global minimization approach followed here,
all molecules are translated and rotated simultaneously from their
original position before a new local optimization is carried out.
Subsequent moves are then performed starting from the current local minimum in a
Markov chain.
This approach was not very efficient for large
PAH assemblies and/or extended individual molecules due to the frequent
intersections induced by the large amplitude motions. The parallel
tempering strategy, which employs smaller physical moves, avoids such
problems but can be quite slow in converging toward low energies. In
addition the set of temperatures needs to be adjusted for the
configurations to be communicated to the colder trajectories. In practice,
we supplemented the parallel tempering MC simulations with periodic
quenches at the trajectory corresponding to $T=20$~K.
Several of the most stable structures were found by seeding the optimizations
with a single main stack and adding the required extra
molecules randomly in space.

The results obtained in the previous section and the occurrence of small stacks
as the main structural motif of larger assemblies suggest several ways for
improving the basin-hopping technique in order to deal more efficiently
with such anisotropic molecules as PAH's:
\begin{itemize}
\item[(i)] Randomly select a single molecule; choose its orientation from
among those of the other molecules; move the molecule randomly in space to a
location where it does not suffer any overlap.
\item[(ii)] Select several molecules stacked together and move then
simultaneously as a block, so that they do not overlap
with any remaining molecule.
\item[(iii)] Follow collective, large amplitude motion with a rescaling of
all distances, such that the minimum pair distance is at least the PAH size;
then choose random orientations for all molecules.
\item[(iv)] Perform local moves that add or remove a single molecule
at the tip of a stack.
\end{itemize}

The above ideas can be combined into a single Monte Carlo approach,
with appropriate probabilities for each type of move. They are suited to
the specific case of prolate molecules, which pose more significant
difficulties in numerical simulations than spherical or oblate species.
Such moves obviously constitute a bias to optimization, but they should
improve the performance of the algorithm by guiding it
toward the multiple-stack funnels.

\subsection{Influence of the potential}

The choice of the intermolecular potential is crucial in predicting
the correct physical and chemical properties of macromolecular assemblies.
This is especially true for smaller molecules, such as benzene, where
small changes in the repulsion-dispersion or electrostatic
parameters can easily lead to the wrong crystal structure.\cite{benzenebulk}

\begin{figure}[htb]
\setlength{\epsfxsize}{6cm}
\leavevmode\centerline{\epsffile{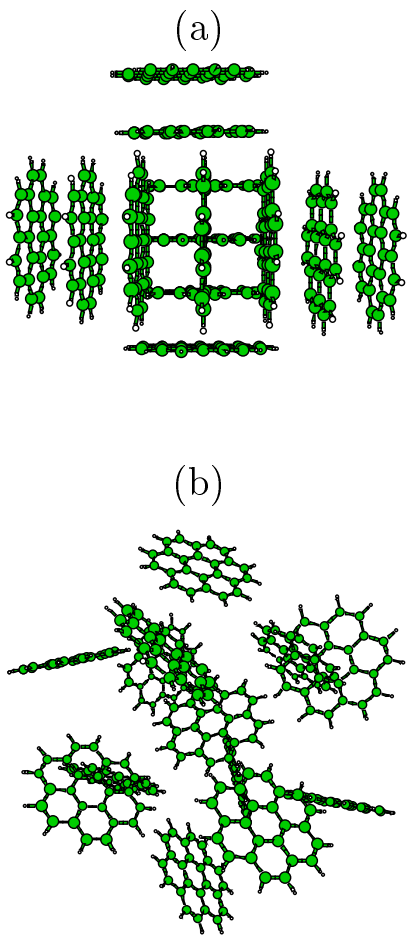}}
\caption{Putative global minima obtained for (coronene)$_{13}$ based on
the same potential energy function, Eq.~(\protect\ref{eq:vpot}) but different
sets of rigid partial charges obtained from 
(a) EPF charges, and (b) NBO charges.}
\label{fig:coro13}
\end{figure}

We did not find significant changes in our results on taking the
Lennard-Jones parameters from other sources, such as the OPLS model
of Jorgensen and coworkers.\cite{opls} However, the same is not true for
the electrostatic multipolar contribution. The NBO charges obtained from
the same DFT calculation on the coronene molecule are significantly
larger in magnitude than the EPF charges, especially on
the hydrogens (Table \ref{tab:charge}).
The corresponding Coulomb energy is therefore
much larger, and becomes comparable to the
dispersion energy. Such charges yield very different global minima,
as illustrated in Fig.~\ref{fig:coro13} for (coronene)$_{13}$.
The disappearance of the handshake structure, Fig.~\ref{fig:coro13}a,
reflects the much longer range character of the intermolecular interaction,
which somewhat shields the molecular anisotropy.
Here it becomes necessary to clearly distinguish between the anisotropy
of the molecules themselves (the so-called shape anisotropy)
and the effective anisotropy of their intermolecular potential.
Increasing the magnitude of the Coulomb interaction results in a stronger
short-range repulsion, which is first manifested in destabilization of the
simple stack, either crossed or parallel-displaced, for
the coronene dimer. At their new equilibrium distance, the relative
orientation between two molecules is mainly driven by the
interaction between quadrupoles. For larger PAH molecules, we expect the
partial charges and quadrupole to become less and less important as the
carbon frame gets more homogeneous. On the contrary, smaller PAH's should
be even more sensitive to the electrostatic effects. This is consistent
with the equilibrium structures, which favour the T-shaped motif
over parallel stacks.

In larger assemblies, isotropic interactions favor an icosahedral
growth scheme over stacking. In the structure displayed in 
Fig.~\ref{fig:coro13}b, the molecular centers of mass indeed form
a distorted icosahedron, similar to what has been found in (benzene)$_{13}$
by several authors.\cite{vandewaal1,bartell,easter1,easter2} Such
structures are significantly easier to locate than stacked isomers,
because the orientational degrees of freedom can be neglected in a
first approach. Here the third trick suggested in the previous
subsection for improving the basin-hopping algorithm should be
efficient.

The Mulliken charges responsible for the icosahedral minimum are probably
much too high and they do not support stable stacks of coronene PAH's, 
which is in disagreement with most previous studies. Conversely, setting
the partial charges to zero does not alter the results very much,
thus confirming that stacking is mostly induced by the intrinsic molecular
anisotropy.

\section{Conclusion}
\label{sec:ccl}

The work reported in this paper was devoted to locating the global
minimum for assemblies of large polycyclic aromatic hydrocarbon
molecules. To do this we first constructed an empirical potential from
available data for the repulsion-dispersion part, as well as from
some carefully analysed density functional calculations for the
electrostatic contribution. These extra {\em ab initio}\/ calculations
were mainly motivated by the lack of published data on the specific
interactions between large PAH molecules. In an attempt to rationalize
the various charge distributions obtained for the neutral and ionic
PAH's, we adjusted a simple fluctuating charges model based on four
parameters. This model, used for reoptimizing the charges of
the PAH molecules in the environment of their own stable assemblies,
indicates that the fixed partial charges approximation is quantitatively
correct, at least close to the equilibrium geometries.

Using a combined set of global optimization techniques, we have shown
that clusters of PAH molecules grow by forming stacks. This
optimization process turned out to be difficult because of the large
amplitude random motions involved. Therefore the putative global minima
we have found might be bettered in future work, although we expect the
trends to be correct. We suggested ways of
improving optimization algorithms dealing with strong shape anisotropy, which
could also be useful for the liquid crystal phases of discotic
molecules.\cite{bates,zewdie,caprion}

Stack formation appears to be mainly driven by the PAH size, since
clusters of benzene and, to some extent, naphthalene or anthracene,
tend to keep some of the polytetrahedral order known in small
atomic clusters.\cite{vandewaal1} However, the one-dimensional growth
of the stack is quite limited, and more close-packed forms become
favored after a certain size, which increases with the PAH diameter.
These new forms are based on smaller stacks, and larger clusters
are also made of the small stack motif. Eventually, parallel stacks
with alternate orientations of the molecular plane appear, as the
precursor to the herringbone crystalline arrangement. We also found
that the electronic structure calculations of the partial charges had to
be performed carefully, since overestimation of the Coulomb energy
wrongly leads to more isotropic interactions, which destabilizes the stacks.

The formation and physico-chemical evolution of PAH's 
is a question of great interest in astrochemistry. Recently, 
Rapacioli {\em et al.}\cite{rapacioli} have revealed the chemical 
link that exists between
PAH molecules and small carbonaceous grains. These authors suggested that 
the grains might be PAH clusters. The results presented here can provide 
guidelines to understand the nature and evolution of such systems 
in astrophysical environments. We note that all these results are only based 
upon energetic considerations and that entropic 
criteria have not been taken into account. Growth could proceed by monomer
addition, but also by the addition of small clusters, presumably as
short stacks. Therefore, the aggregation kinetics could also play an 
important role in the structure of the final aggregate. As we have seen, 
thermodynamics should favor segregation, since
smaller PAH's are expected to be located in the
outermost parts of the aggregate. Upon addition of one or more extra
PAH's, the rearrangements required to reach the true global minimum
involve very high energy barriers, due to the anisotropy
of the molecules and their layered structures. 
In the interior of molecular clouds the temperature is 
very low, which might be an argument against the relevance 
of calculations based on thermodynamics only. However, the large
timescales involved in molecular clouds, whose lifetime is close to
$\sim$10$^6$ years, could allow the cluster to explore many
configurations, eventually reaching low-lying minima.
Another opportunity for rearrangement into these stable
configurations comes from heating
by UV-visible photons originating from stars at the border of the clouds. 
It is in such regions that Rapacioli {\em et al.}\cite{rapacioli} 
have reported photophysical data for PAH clusters,
through their mid-IR emission spectrum
and photoevaporation properties. In these heated regions, we predict that 
for mixed clusters the smallest PAH's are located at the periphery
and are therefore photoevaporated first. If this is the case, the composition 
of both aggregated and free interstellar PAH's is expected to strongly depend 
on UV processing. 

In this paper we also briefly discussed the intermolecular vibrational
modes. Although our results are very preliminary, they 
already give an idea of the frequencies expected for such modes. 
For the stiffest modes, values between 55.6 and 81.3 cm$^{-1}$ are found. 
They could correspond to bands in the astronomical spectra and can 
be searched for in the data obtained from the Long Wavelength Spectrometer, which
was on board the Infrared Space Observatory.\cite{Kessler}
One difficulty
is to disentangle these bands from the numerous intense molecular lines
at moderate resolution.\cite{lorenzetti}  
Hopefully, the PAH cluster features will be revealed by future space
missions in 
the far-IR to submillimeter range, in particular, the Herschel Space 
Observatory will explore the 16--175 cm$^{-1}$ range.

In addition to the intermolecular modes obtained from the present
model, the influence of the cluster structure on the intramolecular frequencies needs
to be quantified, and compared to available astronomical 
spectra.\cite{rapacioli} Investigating the coupling between inter- and
intramolecular modes requires a more realistic atomistic description,
beyond the rigid-body approximation. Extensions of the tight-binding
Hamiltonian developed by the Parneix group\cite{parneix} are currently
under development.

\end{document}